\documentstyle[12pt]{article}
\begin{document}
\begin{center}
{\bf Percolation-like phase transition in a non-equilibrium steady state}
\vskip 1cm
Indrani Bose and Indranath Chaudhuri
\\Department of Physics, \\Bose Institute,
\\93/1, Acharya Prafulla Chandra Road,
\\Calcutta-700 009, India.
\end{center}

\begin {abstract}
We study the Gierer-Meinhardt model of reaction-diffusion on a site-disordered
square lattice. Let $p$ be the site occupation probability of the square 
lattice. For $p$ greater than a critical value $p_c$, the steady state
consists of stripe-like patterns with long-range connectivity. For $p\,<\,p_c$,
the connectivity is lost. The value of $p_c$ is found to be much greater than 
that of the site percolation threshold for the square lattice. In the vicinity
of $p_c$, the cluster-related quantities exhibit power-law scaling behaviour.
The method of finite-size scaling is used to determine the values of the fractal
dimension $d_f$, the ratio, $\frac{\gamma}{\nu}$, of the average cluster size exponent
$\gamma$ and the correlation length exponent $\nu$ and also $\nu$ itself. The values
appear to indicate that the disordered GM model belongs to the universality 
class of ordinary percolation.
\end{abstract}

PACS number(s): 05.70. Ln

\section*{I. Introduction}
Phase transitions in the non-equilibrium are not as well understood as in the
case of equilibrium systems. The percolation model provides an example of
phase transition in disordered systems \cite{Stauffer,Sahimi}. The model is simply defined.
Consider a square lattice of sites. The lattice can be made disordered by
assuming that each site is present with probability $p$ and absent with probability
$1-p$. There is a critical value $p_c$ of $p$, called the percolation threshold,
below which long-range connectivity in the system under study is missing and above which the
connectivity is present with probability one. The percolation transition is
analogous to equilibrium thermodynamic phase transitions and exhibits ``critical"
phenomena in the vicinity of the percolation threshold. Percolation-like transitions
have been observed in the non-equilibrium in a surface-reaction model \cite{Ziff}, an
autocatalytic reaction model \cite{Aukrust} and chemical wave propagation on a lattice
of excitable and non-excitable clusters \cite{I}. Ziff, Gulari and Barshad(ZGB)
\cite{Ziff} have studied a kinetic reaction model describing the chemical reaction
$CO$+$O$\,$\rightarrow$\,$CO_2$ on a catalytic surface. The surface is represented by a square lattice,
the sites of which can be singly occupied by a $CO$ molecule or an $O$ atom.
let $X_{CO}$ and 1-$X_{CO}$ be the molecular concentrations of $CO$ and $O_2$ 
molecules in the gas above the surface. $CO$
and $O_2$ molecules are added randomly to the lattice in the relative ratio of
their molecular concentrations. Two adjacent empty sites are necessary for the
adsorption of an oxygen molecule. The reaction consists of $CO-O$ pairs being
randomly selected from nearest-neighbour (n.n) lattice sites occupied by $O$ 
atoms and $CO$ molecules. The reaction produces $CO_2$ molecules which escape
from the surface. ZGB found that when $X_{CO}$ is less than a certain value $X_1$,
the lattice is completely covered by oxygen atoms in the steady state and no
further reaction takes place. For $X_{CO}$ greater than $X_2$, the lattice is filled
with $CO$. The system exists in a catalytically active mixed-phase only in the
range $X_1\,<\,X_{CO}\,<\,X_2$. The transitions at $X_1$($X_2)$ have been found to be of
second (first) order. Later studies \cite{Dickman,Grinstein} have shown that the second-order
phase transition belongs to the universality class of Reggeon-field theory (RFT)
to which directed percolation (DP) also belongs. Aukrust et al \cite{Aukrust} have
studied an irreversible kinetic reaction model for a one-component autocatalytic
reaction $A+A$\,$\rightarrow$\,$A_2$. In this model, an atom adsorbing on a lattice site reacts,
with probablity 1-$p$, with one of the n.n. atoms, if present. After the reaction,
the two atoms leave the lattice. Otherwise the atom occupies the site. As $p$
is varied, the model undergoes a second-order kinetic phase transition from
a partial occupation (chemically-reactive state) of the lattice to a completely
covered inert state. Again, the phase transition has been shown to belongs
to the universality class of Reggeon-field theory-directed percolation.
Sendi$\tilde{n}$a-Nadal et al \cite{I} have studied the propagation of chemical waves
in a disordered excitable medium in terms of percolation theory. They performed
an experiment with Belousov-Zhabotinsky reagent catalyzed by $Ru(bpy)$ complex
which is sensitive to visible light thus making it possible to control 
the excitability of the system experimentally. The system consists of 
excitable (black) and nonexcitable (white) clusters. The effective wave front 
velocity has been observed to jump from 0 finite values at a threshold $p=p_c$,
where $p$ is the proportion of black sites. At $p_c$, a cluster of black sites
spans the medium. The data obtained appear to indicate that the percolation 
process is similar to ordinary percolation.
In this paper, we study the Gierer-Meinhardt (GM) model \cite{Koch} of reaction-diffusion (RD)
in two dimensions (2d) and show that a percolation-like transition occurs in 
the non-equilibrium steady state as the underlying lattice is made disordered.
In the vicinity of the percolation threshold $p_c$, evidence of power-law 
scaling behaviour of cluster-related quantities is obtained. The percolation
threshold appears to be in the universality class of ordinary percolation.
The study is based on the finite-size scaling analysis of the percolation problem.
In Section II, we define the GM model of RD and describe the various results 
obtained by us. Section III contains concluding remarks.

\section*{II. Reaction-diffusion in the presence of disorder}

Turing \cite{Turing} made a remarkable suggestion, several years back, that 
diffusion need not always act to smooth out concentration differences in
a chemical system.
Two interacting chemicals can generate a stable,  
inhomogeneous pattern if one of the substances (the inhibitor) 
diffuses much faster than the other (the activator). The
activator is autocatalytic, i.e, a small increase in its concentration
`a' over its homogeneous steady-state concentration leads to a further
increse of `a'. The activator besides promoting its own production
also promotes the production of the inhibitor. The inhibitor, as
the name implies, is antagonistic to the activator and inhibits
its production. Suppose, originally the system is in a
homogeneous steady state. A local increase in the activator concentration
leads to a further increase in the concentration of the activator
due to autocatalysis. The concentration of the inhibitor is also increased
locally. The inhibitor, having a diffusion coefficient much
larger than that of the activator, diffuses faster to the
surrounding region and prevents the activator from coming there. This
process of autocatalysis and long-rang inhibition finally lead
to a stationary state consisting of islands of high activator
concentration within a region of high inhibitor
concentration. The islands constitute what is known as the 
Turing pattern. In some Turing systems, there is a possibility of saturation
of the autocatalytic process. In this case, the inhibitor production is limited.
Locally activated regions now have activated neighbours. Nonactivated areas
are, however, close by into which the inhibitor can diffuse. This results in
the emergence of stripe-like patterns which have long-range connectivity.
We study the effect of disorder on the connectivity. The Turing patterns are regions
of chemical concentration gradients. Turing's original idea was that the 
gradients are responsible for the patterns seen in biological systems. Later,
several physical, chemical and biological systems have been identified
\cite{Castets,Ouyang,Lee,Cross,Astrov,Kondo} which exhibit Turing patterns.

The GM model of RD is described by the following two partial differential
equations :

\begin{eqnarray*}
\frac{\partial a}{\partial t}\,&=&\,D_a\,\Delta a\,+\,\rho_a\,\frac{a^2}{(1+k_a\,a^2)h}\,
-\,\mu_a\,a \quad\quad\quad\quad          (1a)      \\
\frac{\partial h}{\partial t}\,&=&\,D_h\,\Delta h\,+\,\rho_h\,{a^2}\,
-\,\mu_h\,h \quad\quad\quad\quad          (1b)     
\end{eqnarray*}
         where $\Delta $ is the Laplacian given by $\Delta \,=\,\frac{\partial^2}
         {\partial x^2}\,+\,\frac{\partial^2}{\partial y^2}\,$, `a' and `h' denote the concentrations 
of the activator and the inhibitor, $D_a,\,\,D_h$, are the respective diffusion
coefficients, $\mu_a,\,\mu_h$ are the removal rates and $\rho_a,\,\rho_h$
are the cross-reaction coefficients. The conditions for the formation of 
stable Turing patterns are $D_h\,>>D_a$ and $\mu_h\,>\mu_a$ \cite{Koch}.
We also assume that $\frac{\rho_a}{1+k_a}\,=\mu_a$ and $\rho_h\,=\mu_h$. In this case,
the steady state solution of equations (1a) and (1b) is given by (a,h)=(1,1),
i.e., the steady state is homogeneous.  
We use the parameter values $D_a\,=0.005$, $D_h\,=0.2$, $\rho_h\,=\mu_h\,=0.02$,
$\rho_a\,=0.0125$ and $k_a\,=0.25$ for our study.
Bose and Chaudhuri \cite{Bose} have studied the effect of disorder on the formation
of Turing islands in the GM model with $k_a\,=0$. We extend this study to the 
case of $k_a\,\neq0$. As in Ref.\cite{Bose}, a very simple discretization scheme
is used. The lattice chosen is of size $L$\,$\times$\,$L$, with $L$ ranging from 20
to 70 in steps of 10. Also, periodic boundary conditions are used. Disorder is 
introduced into the underlying substrate (lattice) by stipulating that the 
probability of a paticular site being part of the RD network is $p$. The 
disordered lattice configuration is generated with the help of a random number 
generator \cite{Bose}. 
For the disordered lattice, the Laplacian is written as

\begin{eqnarray*}
\Delta\,a(x_{ij},\,t)\,&=&\,iocc(i+1,\, j)\,\times\,(\,a(x_{i+1j},\, t)-a(x_{ij},\,t)\,)\\
&+&iocc(i,\,j+1)\,\times\,(\,a(x_{ij+1},\,t)-a(x_{ij},\,t)\,)\\
&+&iocc(i-1,\,j)\,\times\,(\,a(x_{i-1j},\,t)-a(x_{ij},\,t)\,)\\
&+&iocc(i,\,j-1)\,\times\,(a(x_{ij-1},\,t)-a(x_{ij},\,t)\,)  \quad\quad\quad\quad      (2)
\end{eqnarray*}
$x_{ij}$ denotes a lattice site (i,j).
The array $iocc$ keeps track of the occupation status of the sites of the square
lattice. If the site $x_{ij}$ is occupied then $iocc(i,j)$=1, otherwise it is 
equal to zero. Eqn.(2) expresses the fact that diffusion to a site from a 
neighbouring site takes place only if the neighbouring site belongs to the RD
network.  
One can easily check that the discretized
differential equations (with $\frac{\rho_a}{1+{k_a}}\,=\mu_a$ and $\rho_h\,=\,\mu_h$) have a
steady state solution given by a=1 and h=1 for all the lattice sites.
Random fluctuations of magnitude less than 0.1 are created in the steady state with
the help of the random number generator. This fixes the values of a and h at all
the cluster sites at time t=0. The values of a and h at time t+1 are determined
at a site $x_{ij}$ belonging to a cluster with the help of the discrete equations
for a and h. This process is repeated till the steady state is reached, i.e., the
values of a and h at all the cluster sites do not change within a specified accuracy.
We define an `activated' site as one at which the activator concentration has a value  
greater than 1, which is the homogeneous steady-state value. 
Figs. 1(a)-1(c) show the patterns of activated sites on a lattice of size 
50$\times$50 and for site occuptation probability $p=1.0$, $p=0.9$, and $p=0.59$.
Fig. 1(a) shows a fully-connected stripe pattern with long-range connectivity. 
For $p$ greater than a critical value $p_c$, a cluster of n.n. activated sites 
still spans the lattice. Fig. 1(b) shows a pattern in which long-range connectivity
still exists. Fig. 1(c) shows a pattern without long-range connectivity ($p\,<\,p_c$).
The spanning of the largest cluster is checked as in the percolation problem
and the value of $p_c$ is measured by the method of binary chopping \cite{Stauffer1}.
Table I shows the average $p_c$ values determined for lattices of various sizes.
For each lattice size, $p_c$ is calculated by taking the average of 100 values.
Because of the finite size of the lattice, the value of $p_c$ has a small spread.
As in the percolation problem, one expects that $p_c$ has a sharp, unique value
as the lattice size L$\rightarrow$\,$\propto$. The critical value $p_c$ is greater than
the site percolation threshold $p_c^{site}\,=0.59$ for the square lattice.

We now describe some cluster-related properties. Fig. 2 shows the spanning 
cluster for RD on a 50$\times$50 lattice and at $p$=$p_c$ (Table I). Let M be
the size of the spanning cluster, size denotes the number of sites in the cluster.
Size of the largest cluster is determined for lattice size L$\times$L, L ranging
from 20 to 70 in steps 10. To calculate M for each lattice size, the average of 
200 values is taken. Table II shows the data for average M for various lattice
sizes. The fractal dimension $d_f$ of the largest cluster is defined as
\begin{center}
M\,$\sim$\,$L^{d_f}$     \quad\quad\quad\quad       (3)
\end{center}
Fig. 3 shows a plot of $\log(M)$ vs. $\log(L)$. From the slope, the value of $d_f$
is obtained as $d_f$=1.86$\pm$\,0.01 . The error quoted is the least-square fitting error.
The value of $d_f$ is close to that of the spanning cluster ($d_f$=1.89) of 
ordinary percolation.
At any value of $p$$\leq$$p_c$, the non-equilibrium steady-state pattern consists 
of clusters of various sizes. Let $n_s$ be the number of clusters of size $s$
per site of the lattice. The average cluster size $S_{AV}$ is then defined as
\begin{center}
$S_{AV}$\,$\sim$\,$\acute{\sum}_s$ ${s^2}{n_s}$     \quad\quad\quad\quad      (4)
\end{center}
The prime indicates that the largest cluster is not included in the sum. As
$p$\,$\rightarrow$\,$p_c$, $S_{AV}$ exhibits a power-law scaling of the form
\begin{center}
$S_{AV}$\,$\sim$\,$\left|p-p_c\right|^{-\,\gamma}$   \quad\quad\quad\quad    (5)
\end{center}
The critical exponent $\gamma$ can be determined by various numerical methods.
We use the method of finite-size scaling to obtain an estimate of $\gamma$ 
in our model. In this method, the average cluster size as a function of $p$ and
L is given by 
\begin{center}
$S_{AV}(p,L)\,\sim\,L^\frac{\gamma}{\nu}\,f((p-p_c)L^\frac{1}{\nu})$  \quad\quad\quad\quad  (6)
\end{center}
The exponent $\nu$ is the correlation length exponent. As $p$\,$\rightarrow$\,$p_c$, the
correlation length $\xi$ diverges as
\begin{center}
$\xi\,\sim\,\left|p-p_c\right|^{-\,\nu}$  \quad\quad\quad\quad   (7)
\end{center}
The function $f$ in Eqn.(6) is a suitable scaling function and at 
$p$=$p_c$, $f(0)$=constant. The average cluster size is determined for lattice size L 
ranging from 20 to 70 in steps of 10. For each lattice size, the average of 200
values is taken. Table III shows the data obtained for $p$=$p_c$. Fig. 4 shows
a plot of $\log(S_{AV})$ versus $\log(L)$ for $p$=$p_c$. From the slope of the 
straight line, the exponent $\frac{\gamma}{\nu}$ is estimated as $\frac{\gamma}{\nu}=1.72\pm0.01$.
Again, the error quoted is the least-square fitting error. The value of 
$\frac{\gamma}{\nu}=1.79$. The average cluster size $S_{AV}$ is further determined 
for ($p-p_c$) in the range 0.03 to 0.33. Fig. 5 shows a plot of 
$\frac{S_{AV}(p,L)}{L^\frac{\gamma}{\nu}}$ versus z=$(p-p_c)\,L^{\frac{1}{\nu}}$.
Data for three lattice sizes (L=50, 60 and 70) are shown using the symbols:
solid circle (L=50), open circle (L=60) and solid square (L=70). According
to Eqn. (6), all the data should fall on a single curve with the functional 
form $f(z)$. The value of the correlation length exponent $\nu$ for which the
best collapse of data is obtained is $\nu=1.66$. The value of $\nu$ is $\nu=1.75$ in
the case of ordinary percolation.

\section*{III. Conclusion}

We have studied a RD model on the square lattice. The non-equilibrium steady
state of the model consists of stripe-like patterns with long-range connectivity.
A stripe is a region of activator concentration greater than the homogeneous, steady
state value of 1. The sites of the underlying square lattice are occupied
with probability $p$. As $p$ falls below a critical value $p_c$, the long-range
connectivity of the steady state pattern disappears. As $p\,\rightarrow\, p_c$,
the average cluster size diverges. The spanning cluster at $p=p_c$ is a
fractal object. The values of the exponent $\frac{\gamma}{\nu}$ and the fractal 
dimension $d_f$ appears to indicate that the percolation model belongs to the 
universality class of ordinary percolation. The value 1.66 of the correlation
length exponent $\nu$ is different from the value 1.75 in the case of ordinary
percolation. However, the difference may be due to finite-size effects and
also due to the fact that the number of trials ($\sim\,200$) over which an 
average is taken is not large. This is because each trial takes a considerable
amount of time. First, a steady state has to be generated which is reached only
after several thousands of time steps. It is for this very reason that the 
lattice size could not be made large.
The surface reaction models, we have discussed before, belong to the universality
class of directed percolation whereas the GM model of RD appears to belong to 
the universality class of ordinary percolation. In the ordinary percolation
problem, one looks at the connectivity of sites which are randomly occupied with
probability $p$. The occupation status of a particular site is independent
of the occupation status of neighbouring sites. In our study, we focus on the 
connectivity of a RD steady state pattern formed on a disordered lattice.
The sites of the pattern are not randomly occupied, as in the case of ordinary
percolation, but there is some degree of correlation in the occupancy of
sites. The distribution of sites in the steady-state depends on the parameters 
of the RD model, the diffusion coefficients $D_a$, $D_h$, the saturation
parameter $k_a$ etc. The cluster-related properties of the steady state pattern,
say, the average cluster size, have a power-law scaling behaviour when 
occupation probability $p$ of the underlying lattice is close to a critical
value $p_c$. As already mentioned in the Introduction, there are many physical,
chemical and biological systems which exhibit Turing patterns. Some of these
patterns consist of stripes which may sometimes form in a disordered
environment. The present study is of relevance to such systems. In the 
percolation picture, one studies a transition from localised to extended 
states as a function of disorder. In the GM model, instead of disorder, one
can vary the parameter $k_a$ to bring about the transition. For $k_a$=0, the
steady state pattern consists of isolated Turing islands. For $k_a\,\neq\,0$,
the steady state consists of connected stripes. An interesting question to
ask is whether the transition from localised to extended states occurs as 
soon as $k_a$ is made different from zero. Preliminary studies \cite{Chaudhuri}
shows that the transition occurs at a finite value of $k_a$. The connection
of this transition with the one in the presence of disorder will be explored
in future. The effect of changing the parameters of the RD model and also 
changing the form of the nonlinearity in the RD model (provided it still gives
rise to connected patterns in the steady state) will also be investigated.
\section*{Acknowledgment}
One of the author I.C. is supported by the Council of Scientific and Industrial
Research, India under Sanction No. 9/15(173)/96-EMR-I.
\newpage
\begin{center}Table I. The average value of $p_c$ for lattices of various sizes.
\vskip 0.1cm
\begin{tabular}{|c|c|}\hline
Latice size    & Average $p_c$\\ \hline
$20\times 20$   & 0.81089\\
$30\times 30$   & 0.80735\\
$40\times 40$   & 0.80992\\
$50\times 50$ & 0.80714\\
$60\times 60$ & 0.80873\\  
$70 \times 70$& 0.80843\\ \hline

\end{tabular}
\end{center}

\begin{center}Table II. The average  size M of the spanning cluster at $p=p_c$
for lattices of various sizes
\vskip 0.1cm
\begin{tabular}{|c|c|}\hline
Latice size    & Average M\\ \hline
$20\times 20$   & 96.54\\
$30\times 30$   & 195.77\\
$40\times 40$   & 350.65\\
$50\times 50$ &   508.77\\
$60\times 60$ &   754.81\\ 
$70 \times 70$&  974.22\\ \hline

\end{tabular}
\end{center}

\begin{center}Table III. The average  cluster size $S_{AV}$ at $p=p_c$
for lattices of various sizes
\vskip 0.1cm
\begin{tabular}{|c|c|}\hline
Latice size    & Average $S_{AV}$\\ \hline
$20\times 20$   & 4.01\\
$30\times 30$   & 8.22\\
$40\times 40$   & 14.11\\
$50\times 50$ &   19.65\\
$60\times 60$ &   27.82\\ 
$70 \times 70$&  33.96\\ \hline

\end{tabular}
\end{center}

\newpage
\section*{Figure Captions}
\begin{description}
\item[Fig.1] Non-equlibrium steady state patterns in the GM model 
on a disordered square lattice of size $50\times 50$. The probability that a site of the original
square lattice exists is $p$. The patterns are shown for (a) $p$=1, (b) $p$=0.9,
(c) $p$=0.59 .
\item[Fig.2] The spanning cluster on a $50\times 50$ lattice for $p=p_c$ (see Table I.) .
\item[Fig.3] $\log(M)$ versus $\log(L)$ for $p=p_c$ (Eqn.(3)). The slope of the 
straight line gives the fractal dimension $d_f$=1.86$\pm$0.01 .
\item[Fig.4] $\log(S_{AV})$ versus $\log(L)$ for $p=p_c$ (Eqn.(6)). The slope of the
straight line gives the exponent $\frac{\gamma}{\nu}$=1.72$\pm$0.01 .
\item[Fig.5] The collapse of data for three different lattice sizes: solid circles,
open circles and solid squares for L=50, 60 and 70 respectively.
\end{description}

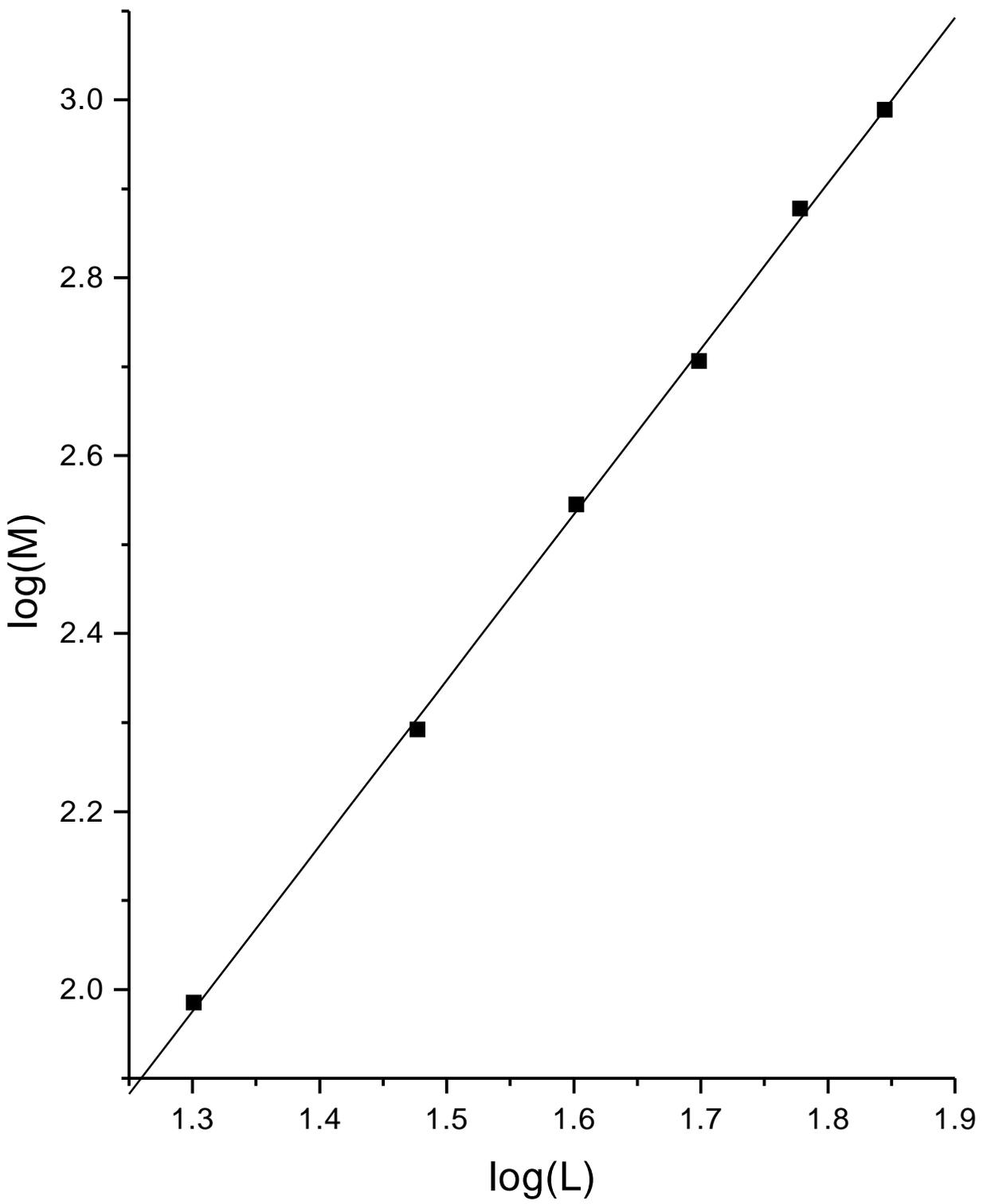
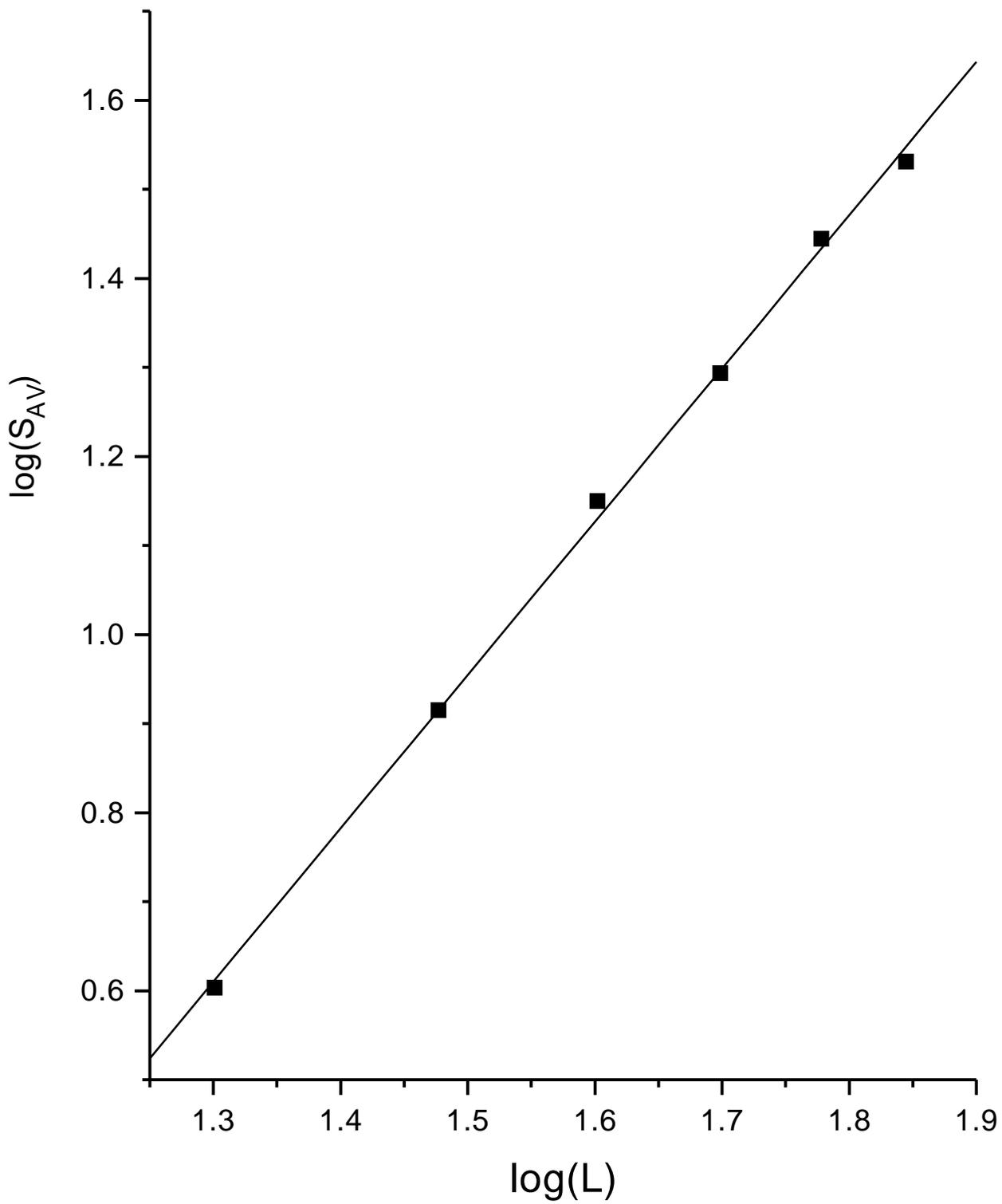
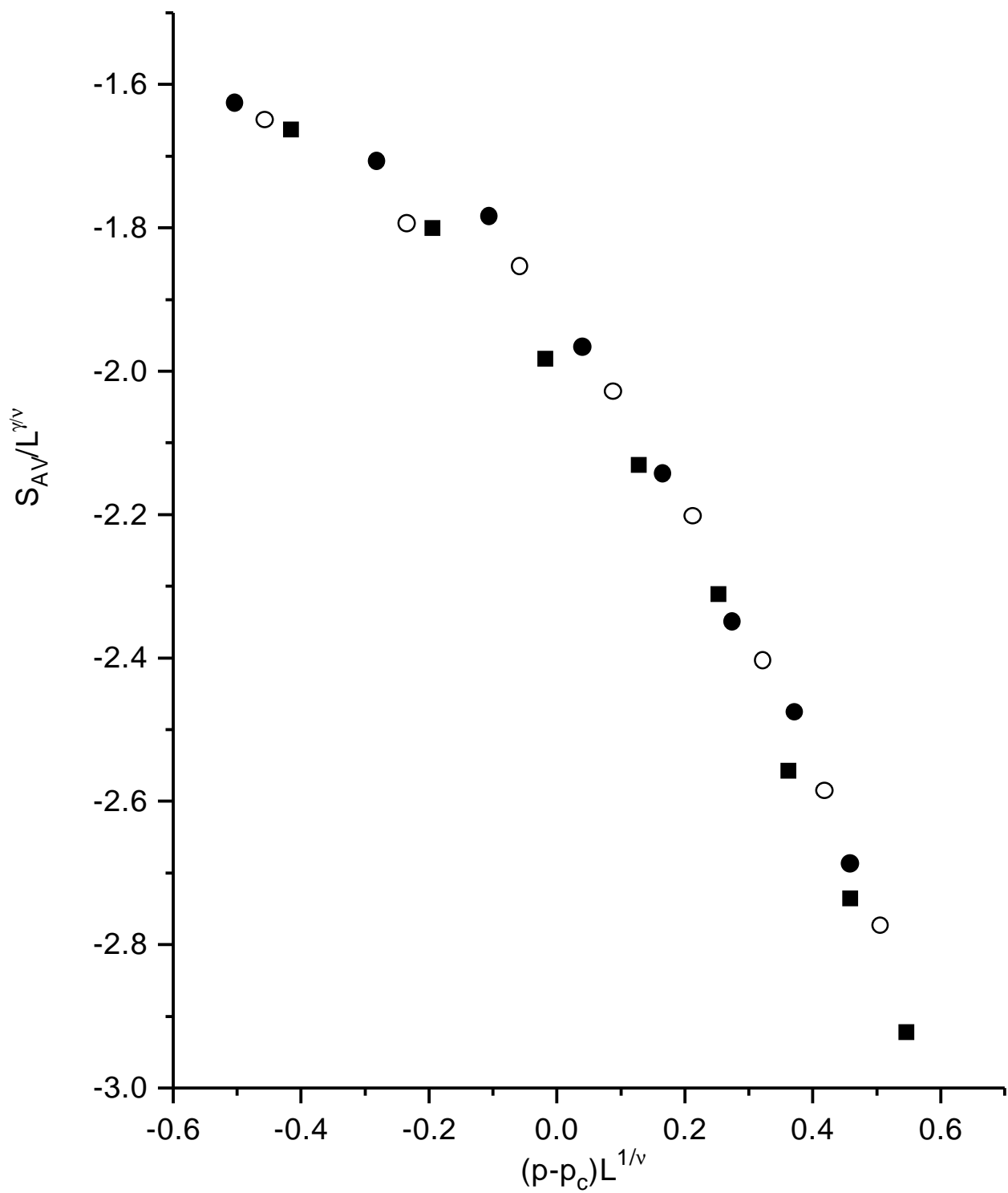
\end{document}